\documentstyle[12pt,psfig]{article}
\newcommand{\EQ}{\begin{equation}}
\newcommand{\EN}{\end{equation}}
\newcommand{\bea}{\begin{eqnarray}}
\newcommand{\eea}{\end{eqnarray}}

\begin{document}
\topmargin 0pt
\oddsidemargin 5mm
\renewcommand{\thefootnote}{\arabic{footnote}}
\newpage
\setcounter{page}{0}
\begin{titlepage}
\begin{flushright}
SISSA REF 62/2002/FM
\end{flushright}
\vspace{0.5cm}
\begin{center}
{\large {\bf Correlation functions of disorder operators} \\
{\bf in massive ghost theories}}\\
\vspace{1.8cm}
{\large G. Delfino$^{1,2}$, P. Mosconi$^{1,3}$ and G. Mussardo$^{1,2}$} \\
\vspace{0.5cm} {\em $^{1}$International School for Advanced
Studies, via Beirut 2-4, 34014 Trieste, Italy }\\
\vspace{0.3cm} {\em $^{2}$Istituto Nazionale di Fisica Nucleare,
Sezione di Trieste} \\
\vspace{0.3cm} {\em $^{3}$Istituto Nazionale
di Fisica della Materia, Sezione di Trieste}\\
\end{center}
\vspace{1.2cm}

\renewcommand{\thefootnote}{\arabic{footnote}}
\setcounter{footnote}{0}

\begin{abstract}
\noindent
The two-dimensional ghost systems with negative integral central charge 
received much attention in the last years for their role in a number of 
applications and in connection with logarithmic conformal field theory. 
We consider the free massive bosonic and fermionic ghost systems and 
concentrate on the non-trivial sectors containing the disorder operators. 
A unified analysis of the correlation functions of such operators can be 
performed for ghosts and ordinary complex bosons and fermions.
It turns out that these correlators depend only on the statistics although the 
scaling dimensions of the disorder operators change when going from the 
ordinary to the ghost case. As known from the study of the ordinary case, 
the bosonic and fermionic correlation functions
are the inverse of each other and are exactly expressible through the 
solution of a non-linear differential equation.
\end{abstract}

\end{titlepage}

\newpage
{\bf 1.}\,\,
Ghost fields, namely the quantum fields violating the usual relation 
between spin and statistics, are very popular in physics since when 
Faddeev and Popov showed their role in the quantisation of non-abelian 
gauge theories. In two dimensions, they have been the object of increasing
interest over the last decade because of their applications in the study of
disordered systems, quantum Hall states, polymer physics and dynamical models 
(see e.g. \cite{Bernard,MR,Saleur,Ruelle}). 

In the massless limit, the fermionic (anticommuting scalars) and bosonic 
(commuting spinors) ghost systems entering the study of these two-dimensional
problems are particularly simple
(free) examples of the vast class of `non-unitary' conformal field theories 
which includes in particular all the conformal theories with negative central 
charge. The central charges of the fermionic and bosonic ghosts are 
$c=-2$ and $c=-1$, respectively \cite{FMS}, and differ only for the 
sign from the central charges of their counterparts with the `right' 
statistics, respectively the commuting complex scalar field and the 
anticommuting complex spinor field. Detailed studies of the $c=-2$ and 
$c=-1$ ghost conformal field theories can be found in \cite{Kausch} and 
\cite{LMRS}, respectively. A comparison between the ghost systems 
and their counterparts with positive central charge has been performed 
in \cite{GL}. These models have also provided a privileged playground for 
logarithmic conformal field theory \cite{Gurarie,Flohr}.

In this note we consider the free {\it massive} bosonic and fermionic ghost 
systems, our interest focusing on the non-trivial sectors of these 
models containing the `disorder' operators which are non-local with respect 
to the ghost fields. We recall that two operators $A(x)$ and $B(y)$ are 
said to be mutually non-local with non-locality phase $e^{2i\pi\alpha}$ if 
the correlation functions containing these operators pick up such a phase
when $A(x)$ is taken once around $B(x)$ on the Euclidean plane.
The presence of a continous spectrum of such disorder operators in the ghost
systems is expected on the same physical grounds discussed in \cite{DGM}
for the ordinary complex bosons and fermions. As a matter of fact, it turns out
that, similarly to what observed at the conformal level in \cite{GL}, the 
ghost systems and the ordinary bosons and fermions are intimately related
also in the free massive case. Actually, it is possible to deal in a compact 
form with the four cases in terms of the two parameters
\EQ
S=\left\{
\begin{array}{cl}
1 & \mbox{for bosons} \\
-1 & \mbox{for fermions} \\
\end{array}
\right.
\EN

\EQ
\varepsilon=\left\{
\begin{array}{cl}
1 & \mbox{for ordinary fields} \\
-1 & \mbox{for ghosts}\,. \\
\end{array}
\right.
\EN
In all cases the mass spectrum consists of a doublet of free particles $A$ and 
$\bar{A}$ with mass $m$. Then, denoting $\Phi_\alpha(x)$ the disorder 
operator exhibiting a non-locality phase $e^{2i\pi\alpha}$ ($e^{-2i\pi\alpha}$)
with respect to (the field which interpolates) the particle $A$ ($\bar{A}$), 
we will show that\footnote{We will use the notation 
$\tilde{\Phi}(x)\equiv\Phi(x)/\langle\Phi\rangle$ throughout this note.} 
\EQ
\langle\tilde{\Phi}_\alpha(x)\tilde{\Phi}_{\alpha'}(0)\rangle=
e^{S\,\Upsilon_{\alpha,\alpha'}(m|x|)}\,,
\label{main}
\EN
where $\Upsilon_{\alpha,\alpha'}(t)$ is a function expressed in terms of 
the solution of a non-linear differential equation of Painlev\'e type.
The main point to be remarked in (\ref{main}) is that the r.h.s. depends
on $S$ but not on $\varepsilon$, which implies that the correlation functions 
of the disorder operators in the bosonic and fermionic ghost systems coincide 
with those for the ordinary complex bosons and fermions, respectively. 
The latter correlators and their inversion property according to the 
statistics were discussed in \cite{DGM,SMJ,BL}. 

The $\varepsilon$-independence of the r.h.s. of (\ref{main}) has to be 
contrasted with the fact that the nature of the operators on the l.h.s. 
does depend on $\varepsilon$. Indeed, the values of the scaling 
dimensions $X_\alpha$ of the operators $\Phi_\alpha$ and of the central 
charge in the ultraviolet limit can be written as
\EQ
c=2^{\delta_{S,\varepsilon}}\varepsilon\,,\hspace{1cm}
X_\alpha=S\alpha(\delta_{S,\varepsilon}-\,\alpha)\,.
\label{cx}
\EN
We now turn to explaining the origin of these results.

\vspace{.3cm}
{\bf 2.}\,\,
We work within the form factor approach in which the correlation functions 
are expressed as spectral series over intermediate multiparticle states after
the computation of the form factors
\EQ
f^\alpha_n(\theta_1,\dots,\theta_n,\beta_1,\dots,\beta_n)=
\langle 0|\tilde{\Phi}_\alpha(0)|A(\theta_1)\dots
A(\theta_n)\bar{A}(\beta_1)\dots\bar{A}(\beta_n)\rangle\,.
\label{ff}
\EN
Here rapidity variables are used to parameterise the energy-momentum of a 
particle as $(e,p)=(m\cosh\theta,m\sinh\theta)$. The form factors can be 
determined in integrable quantum field theories solving a set of functional
equations which in the standard cases (see e.g. \cite{YZ}) require as 
input the exact $S$-matrix (quite trivial in the free case we are dealing 
with) and the non-locality phases between the operators and the particles. 
Clearly, what we need for our present purposes is to understand how to 
modify these equations in order to distinguish the ghost case from that of
ordinary particles discussed in \cite{DGM}. 

The Lagrangians of the free theories we are considering contain a kinetic 
and a mass term, each of them linear in the fields which interpolate the 
particles $A$ and $\bar{A}$. In the case of ordinary spin-statistics, 
hermitian conjugation interchanges these two fields leaving the 
Lagrangian invariant. In the ghost case the operation made in the same way 
would change the sign of the Lagrangian because the terms are reordered  
with the `wrong' statistics. Hence, a real Lagrangian requires that the two 
ghost fields are not exactly the hermitian conjugate of each other. A 
suitable choice of the conjugation matrix for all cases is 
\EQ
C=\left(
\begin{array}{cc}
0 & 1  \\
\varepsilon  & 0 \\
\end{array}
\right)\,.
\EN

We are now in the position to write the form factor equations which read
\bea
f_n^\alpha(\theta_1,\dots ,\theta_i ,\theta_{i+1},\dots,\theta_n,\beta_1,
\dots,\beta_n) & = &
S\,f_n^\alpha(\theta_1,\dots,\theta_{i+1},\theta_i,\dots,\theta_n,\beta_1,
\dots,\beta_n), 
\label{ff1}\\
f_n^\alpha(\theta_1+2i\pi,\theta_2,\dots,\theta_n,\beta_1,\dots,\beta_n) &=& 
\varepsilon\,S\,e^{ 2 i \pi \alpha} 
f_n^\alpha(\theta_1, \dots,\theta_n,\beta_1, \dots,\beta_n), 
\label{ff2}\\
\textrm{Res}_{\theta_1-\beta_1= i \pi} f_n^\alpha(\theta_1, \dots,\theta_n,
\beta_1, \dots,\beta_n)
& = & iS^{n-1}(1-e^{ 2 i \pi \alpha})f_{n-1}^\alpha(\theta_2,
..,\theta_n,\beta_2,..,\beta_n).
\label{ff3}
\eea
We work with $0<\alpha<1$.
We see that only the presence of the factor $\varepsilon$ in the second 
equation distinguish between ghosts and ordinary particles. 
The origin of this factor is the following. Shifting the rapidity of a 
particle by $i\pi$ means inverting the sign of its energy and momentum. 
This inversion, together with charge conjugation, amounts to crossing the 
particle from the initial to the final state. Hence, the $2i\pi$ analytic
continuation in Eq.\,(\ref{ff2}) corresponds to a double crossing from the 
initial to the final state and then again to the initial state, a process
which produces the factor $C^2=\varepsilon$.

The solution to the above equations can be written as
\EQ
f^\alpha_n(\theta_1,\dots,\theta_n,\beta_1,\dots,\beta_n)=
(-i)^{n\,\delta_{S,-\varepsilon}}S^{n(n-1)/2}(-\sin\pi\alpha)^n\,
e^{\left(\alpha-\frac12\delta_{S,\varepsilon}\right)
\sum_{i=1}^n(\theta_i-\beta_i)}\,
\left|A_n\right|_{(S)}\,,
\label{solution}
\EN
where $A_n$ is a $n\times n$ matrix ($A_0\equiv 1$) with entries
\EQ
A_{ij}=\frac{1}{\cosh\frac{\theta_i-\beta_j}{2}}\,,
\EN
and $\left|A_n\right|_{(S)}$ denotes the 
permanent\footnote{The permanent of a matrix differs from the 
determinant by the omission of the alternating sign factors $(-1)^{i+j}$.} 
of $A_n$ for $S=1$ and the determinant of $A_n$ for $S=-1$.

Correlation functions are obtained inserting in between the operators a 
resolution of the identity in the form
\EQ
1=\sum_{n=0}^{\infty}
\int_{-\infty}^{+\infty}\frac{d\theta_{1}...d\beta_{n}}{(n!)^{2}(2\pi)^{2n}}
|A(\theta_1)\ldots A(\theta_n)\bar{A}(\beta_1)\ldots\bar{A}(\beta_n)\rangle 
\langle\bar{A}(\beta_{n})\ldots\bar{A}(\beta_1)A(\theta_n)\ldots A(\theta_1)|.
\EN
Since by crossing and Lorentz invariance we have
\bea
\langle\bar{A}(\beta_{n})\ldots\bar{A}(\beta_1)A(\theta_n)\ldots A(\theta_1)|
\tilde{\Phi}_\alpha(0)|0\rangle &=&
\varepsilon^n\,f_n^{\alpha}(\beta_n+i\pi,\dots,\beta_1+i\pi,
\theta_n+i\pi,\dots,\theta_1+i\pi)\nonumber\\
&=& \varepsilon^n\,f_n^{\alpha}(\beta_n,\dots,\beta_1,
\theta_n,\dots,\theta_1),
\eea
the two-point functions take the form
\EQ
G_{\alpha,\alpha'}^{(S,\varepsilon)}(t)=
\langle\tilde{\Phi}_\alpha(x)\tilde{\Phi}_{\alpha'}(0)\rangle=
\sum_{n=0}^\infty\frac{\varepsilon^n}{(n!)^2\,(2\pi)^{2n}}  
\int d\theta_1\dots d\theta_{n}d\beta_1\dots d\beta_{n} 
\,g_n^{(\alpha,\alpha')}(t\,|\theta_1,\dots,\beta_n)\,,
\label{corr}
\EN
where
\bea
g^{(\alpha,\alpha')}_n (t\,|\theta_1,\dots,\beta_n) &=& 
f_n^\alpha(\theta_1,\dots,\beta_n)f_n^{\alpha'}(\beta_n, \dots,\theta_1)\,
e^{-t e_n} \nonumber\\
&=& (\varepsilon S\sin\pi\alpha\,\sin\pi\alpha')^n\,
e^{(\alpha-\alpha')\sum_{i=1}^n(\theta_i-\beta_i)}\,
\left|A_n\right|_{(S)}^2 \,e^{-t e_n}\,,
\label{g}
\eea
\[
t=m|x|\,\,,\hspace{1cm}e_n=\sum_{k=1}^n (\cosh\theta_k + \cosh\beta_k)\,.
\]
Hence the anticipated $\varepsilon$-independence of these correlators
immediately follows from the cancellation between the factor $\varepsilon^n$ 
contained in $g^{(\alpha,\alpha')}_n$ and the one explicitely appearing in 
(\ref{corr}):
\EQ
G_{\alpha,\alpha'}^{(S,\varepsilon)}(t)=G_{\alpha,\alpha'}^{(S)}(t)\,.
\EN
Without repeating the discussion of Ref.\,\cite{DGM}, we simply recall that 
the spectral series for $G_{\alpha,\alpha'}^{(S)}(t)$ can be resummed in
a Fredholm determinant form making transparent the result (\ref{main}),
namely that the bosonic and fermionic correlators are the inverse of each 
other. The function $\Upsilon_{\alpha,\alpha'}(t)$ is given by
\cite{SMJ,BL} 
\EQ
\Upsilon_{\alpha,\alpha'}(t)=\frac12\int_{t/2}^\infty\rho d\rho\,
\left[(\partial_\rho\chi)^2-4\sinh^2\chi-
\frac{(\alpha-\alpha')^2}{\rho^2}\tanh\chi\right]\,,
\label{Sigma}
\EN
where $\chi(\rho)$ satisfies the differential equation
\EQ
\partial^2_\rho\chi+\frac{1}{\rho}\,\partial_\rho\chi=2\sinh 2\chi+
\frac{(\alpha-\alpha')^2}{\rho^2}\,\tanh\chi\,(1-\tanh^2\chi)\,,
\label{diff}
\EN
subject to asymptotic conditions such that for $\alpha+\alpha'<1$ one obtains
\EQ
\lim_{t\rightarrow 0}G_{\alpha,\alpha'}^{(S)}(t)=\left(C_{\alpha,\alpha'}\,
t^{2\alpha\alpha'}\right)^{-S}\,.
\label{uv}
\EN
The amplitude follows from the work of Ref.\,\cite{russians} and reads
\EQ
C_{\alpha,\alpha'}=2^{-2\alpha\alpha'}\exp\left\{2\int_0^\infty\frac{dt}{t}
\left[\frac{\sinh\alpha t\,\cosh(\alpha+\alpha')t\,\sinh\alpha't}{\sinh^2t}
-\alpha\alpha'e^{-2t}\right]
\right\}\,.
\label{ampl}
\EN

\vspace{.3cm}
{\bf 3.}\,\,
The central charge of the ultraviolet limit and the scaling dimensions of 
the operators can be obtained in our off-critical framework through the 
sum rules \cite{cth,DSC}
\EQ
c=\frac{3}{4\pi}\int d^2x\,|x|^2 \,\langle\Theta(x)\Theta(0)
\rangle_{connected}\,,
\EN
\EQ
X_\alpha= -\frac{1}{2\pi}\int d^2x\,
\langle\Theta(x)\tilde{\Phi}_\alpha(0)\rangle_{connected}\,,
\EN
where $\Theta(x)$ denotes the trace of the energy-momentum tensor. Since the 
only non-zero form factor of this operator in the free theories we are dealing 
with is
\EQ
\langle 0|\Theta(0)|A(\theta)\bar{A}(\beta)\rangle=2\pi m^2\left[-i
\sinh\frac{\theta-\beta}{2}\right]^{\delta_{S,-\varepsilon}}\,,
\EN
it easy to check that the sum rules yield the results (\ref{cx}).

For the discussion of the short distance behaviour of the correlators
define the exponent $\Gamma_{\alpha,\alpha'}$ through the relation
\EQ
\langle\Phi_\alpha(x)\Phi_{\alpha'}(0)\rangle\sim |x|^{-
\Gamma_{\alpha,\alpha'}}\,,
\hspace{1cm}|x|\rightarrow 0\,.
\EN
The result (\ref{uv}) for $\alpha+\alpha'<1$ follows from the 
operator product expansion 
\EQ
\langle\Phi_\alpha(x)\Phi_{\alpha'}(0)\rangle\sim |x|^{
X_{\alpha+\alpha'}-X_{\alpha}-X_{\alpha'}}
\langle\Phi_{\alpha+\alpha'}\rangle+\dots\,.
\label{ope}
\EN
The $\varepsilon$--dependence of the scaling dimensions in (\ref{cx}) affects
only the term linear in $\alpha$ and cancels out in the above combination
leaving
\EQ
\Gamma_{\alpha,\alpha'}=2S\,\alpha\alpha'\,,\hspace{1cm}0<\alpha+\alpha'<1\,.
\EN

It seems more difficult to give a unified description for the range 
$1<\alpha+\alpha'<2$. On the basis of the discussion of Ref.\,\cite{DGM}
we expect that for $S=\varepsilon$ the short distance behaviour (\ref{ope})
still holds provided $\alpha+\alpha'$ is taken modulo $1$. Then one finds
\EQ
\Gamma_{\alpha,\alpha'}=2S\,[\alpha\alpha'+1-(\alpha+\alpha')]\,, 
\hspace{1cm}1<\alpha+\alpha'<2\,.
\label{correzione}
\EN
This result is recovered in the case $S=-1$, $\varepsilon=1$ due to the fact
that the first order off-critical correction becomes leading in this range of 
$\alpha+\alpha'$ \cite{DGM}. The mechanism that should lead to 
(\ref{correzione}) in the remaining case of the bosonic ghost is not clear to 
us at present.

At the border value $\alpha+\alpha'=1$ the correlators develop a logarithmic
correction that is most easily evaluated for the well studied case of ordinary
complex fermions \cite{russians}. One concludes
\EQ
\lim_{t\rightarrow 0}G_{\alpha,1-\alpha}^{(S)}(t)=[{\cal B}_{\alpha}\,
t^{2\alpha(1-\alpha)}\ln(1/t)]^{-S}\,,
\label{log}
\EN
with 
\EQ
{\cal B}_{\alpha}=2^{1-2\alpha(1-\alpha)}e^{-(I_\alpha+I_{1-\alpha})}\,,
\EN
\EQ
I_\alpha=\int_0^\infty\frac{dt}{t}\,\left(\frac{\sinh^2\alpha t}{\sinh^2t}-
\alpha^2e^{-2t}\right)\,.
\EN

An interesting check of our results for the ghost correlation functions
can be performed for the operator $\Phi_{1/2}$ in the fermionic ghost theory.
In fact, the free massive fermionic ghost can formally 
be regarded as a limit of the $\varphi_{1,3}$ perturbation of the minimal 
conformal models with central charge \cite{BPZ}
\EQ
c=1-\frac{6}{p(p+1)}\,,
\label{minimal}
\EN
possessing the spectrum of scalar primary fields $\varphi_{l,k}$ with scaling 
dimensions
\EQ
X_{l,k}=\frac{((p+1)l-pk)^2-1}{2p(p+1)}\,.
\EN
The required values $c=-2$ and $X_{1,3}=0$ are found as\footnote{The 
genuine minimal models of the series (\ref{minimal}) have $p=3,4,\ldots$\,.
It is known, however, that many results can be extended to continous values 
of $p$.} $p\rightarrow 1$. Our operator $\Phi_{1/2}$ with scaling dimension 
$-1/4$ is identified with $\varphi_{1,2}$. From the operator product expansion
of the $\varphi_{l,k}$ we have for $p\rightarrow 1$
\bea
\langle \tilde{\varphi}_{1,2}(x)\tilde{\varphi}_{1,2}(0)\rangle &\simeq &
\frac{|x|^{-2 X_{1,2}}}{\langle\varphi_{1,2}\rangle^2}\left(
1+C\,\langle\varphi_{1,3}\rangle\,|x|^{X_{1,3}}\right)\nonumber \\
&\simeq & \frac{|x|^{1/2}}{\langle\varphi_{1,2}\rangle^2}
\left\{1+C\,\langle\varphi_{1,3}\rangle\,\left[1+(p-1)\ln |x|\right]\right\}\,.
\eea
It can be checked from the known values of the structure constant $C$ 
\cite{DF} and of the vacuum expectation values in $\varphi_{1,3}$--perturbed
minimal models \cite{russians} that $C\,\langle\varphi_{1,3}\rangle=-1$ and
$(p-1)/\langle\varphi_{1,2}\rangle^2={\cal B}_{1/2}\,m^{1/2}$ as 
$p\rightarrow 1$, so that the result (\ref{log}) with $S=-1$ and $\alpha=1/2$ 
is indeed recovered.

\vspace{1cm}
{\bf Acknowledgements.} We thank S. Bertolini for interesting discussions.


\end{document}